\documentclass{acm_proc_article-sp}

\usepackage{booktabs}
\newcommand{\superscript}[1]{\ensuremath{^{\textrm{#1}}}}
\def\sharedaffiliation{\end{tabular}\newline\begin{tabular}{c}}
\def\wu{\superscript{*}}
\def\wg{\superscript{\dag}}
\def\wn{\superscript{\ddag}}

\newfont{\mycrnotice}{ptmr8t at 7pt}
\newfont{\myconfname}{ptmri8t at 7pt}
\let\crnotice\mycrnotice%
\let\confname\myconfname%

\begin{document}

% --- Author Metadata here ---
\toappear{\the\boilerplate\par
{\confname{\the\conf}} \the\confinfo\par \the\copyrightetc}
\permission{Permission to make digital or hard copies of all or part of this work for personal or classroom use is granted without fee provided that copies are not made or distributed for profit or commercial advantage and that copies bear this notice and the full citation on the first page. Copyrights for components of this work owned by others than ACM must be honored. Abstracting with credit is permitted. To copy otherwise, or republish, to post on servers or to redistribute to lists, requires prior specific permission and/or a fee. Request permissions from Permissions@acm.org. }
\conferenceinfo{SocialCom '14,}{August 04 - 07 2014, Beijing, China}
\copyrightetc{Copyright 2014 ACM \the\acmcopyr}
\crdata{978-1-4503-2888-3/14/08 ...\$15.00.\\
http://dx.doi.org/10.1145/2639968.2640057}
% --- End of Author Metadata ---

\title{The Role of Peer Influence in Churn in Wireless Networks}
%
% You need the command \numberofauthors to handle the 'placement
% and alignment' of the authors beneath the title.
%
% For aesthetic reasons, we recommend 'three authors at a time'
% i.e. three 'name/affiliation blocks' be placed beneath the title.
%
% NOTE: You are NOT restricted in how many 'rows' of
% "name/affiliations" may appear. We just ask that you restrict
% the number of 'columns' to three.
%
% Because of the available 'opening page real-estate'
% we ask you to refrain from putting more than six authors
% (two rows with three columns) beneath the article title.
% More than six makes the first-page appear very cluttered indeed.
%
% Use the \alignauthor commands to handle the names
% and affiliations for an 'aesthetic maximum' of six authors.
% Add names, affiliations, addresses for
% the seventh etc. author(s) as the argument for the
% \additionalauthors command.
% These 'additional authors' will be output/set for you
% without further effort on your part as the last section in
% the body of your article BEFORE References or any Appendices.

\numberofauthors{2} %  in this sample file, there are a *total*
% of EIGHT authors. SIX appear on the 'first-page' (for formatting
% reasons) and the remaining two appear in the \additionalauthors section.
%
\author{
% You can go ahead and credit any number of authors here,
% e.g. one 'row of three' or two rows (consisting of one row of three
% and a second row of one, two or three).
%
% The command \alignauthor (no curly braces needed) should
% precede each author name, affiliation/snail-mail address and
% e-mail address. Additionally, tag each line of
% affiliation/address with \affaddr, and tag the
% e-mail address with \email.
%
% 1st. author
%\alignauthor
% 1st. author
\alignauthor
Qiwei Han\wu\wg \\
\email{qiweih@cmu.edu}
% 2nd. author
\alignauthor
Pedro Ferreira\wu\wn\\
\email{pedrof@cmu.edu}       
  \sharedaffiliation
  \begin{tabular}{cccccc}
    \affaddr{{\wu}Department of Engineering and Public Policy{\ }} & & \affaddr{{\wn}Heinz College{\ }} & &  \affaddr{{\wg}Instituto Superior T\'{e}cnico{\ }}\\
    \affaddr{Carnegie Mellon University}            & & \affaddr{Carnegie Mellon University} & & \affaddr{ University of Lisbon}  \\
    \affaddr{Pittsburgh, PA 15213}            & & \affaddr{Pittsburgh, PA 15213}  & & \affaddr{Lisbon, 1049-001}\\
    \affaddr{United States}                  & & \affaddr{United States}   & & \affaddr{Portugal}\\
  \end{tabular}   
}
% There's nothing stopping you putting the seventh, eighth, etc.
% author on the opening page (as the 'third row') but we ask,
% for aesthetic reasons that you place these 'additional authors'
% in the \additional authors block, viz.
% Just remember to make sure that the TOTAL number of authors
% is the number that will appear on the first page PLUS the
% number that will appear in the \additionalauthors section.

\maketitle
\begin{abstract}
Subscriber churn remains a top challenge for wireless carriers. These carriers need to understand the determinants of churn to confidently apply effective retention strategies to ensure their profitability and growth. In this paper, we look at the effect of peer influence on churn and we try to disentangle it from other effects that drive simultaneous churn across friends but that do not relate to peer influence. We analyze a random sample of roughly 10 thousand subscribers from large dataset from a major wireless carrier over a period of 10 months. We apply survival models and generalized propensity score to identify the role of peer influence. We show that the propensity to churn increases when friends do and that it increases more when many strong friends churn. Therefore, our results suggest that churn managers should consider strategies aimed at preventing group churn. We also show that survival models fail to disentangle homophily from peer influence over-estimating the effect of peer influence. 
\end{abstract}

% A category with the (minimum) three required fields
%\category{H.4}{Information Systems Applications}{Miscellaneous}
%A category including the fourth, optional field follows...
%\category{D.2.8}{Software Engineering}{Metrics}[complexity measures, performance measures]

%\terms{Economics, Human Factors}

%\keywords{Peer influence, Homophily, Churn, Wireless Networks} 

\section{Introduction}
Over the past two decades, wireless telecommunications markets experienced rapid change and strong technological growth from 2G to 4G. According to \cite{itu2013}, 6.8 billion mobile subscriptions worldwide saturate the wireless market. In parallel, deregulation opened up markets to multiple entrants supporting both competition and technological innovation. Consequently, carriers need to invest heavily in upgrading their networks to provide quality communications and novel services as well as to ensure that they healthily profit from existing subscribers. However, subscribers have many providers to choose from and can ever more easily transfer from one provider to another as more information about products and services abounds and switching costs reduce \cite{daegon2012}. 

Churn rates measure the proportion of subscribers discontinuing service during a certain period of time. As reported by wireless carriers across the world, average monthly churn rates vary between 1.5\% and 5\% \cite{wei2002,kim2004, fccreport2009}. In other words, wireless carriers can lose 20\% of their subscriber base every year, which poses significant challenges for profitability and growth. Subscriber churn may represent significant economic loss. This loss can be estimated by multiplying the average cost to acquire a new subscriber by the number of subscribers that churn. With average acquisition costs reaching \$600 per subscriber churn may cost the wireless industry billions of dollars every year \cite{berson2000}. On the other hand, keeping an existing subscriber is generally much cheaper and easier. \cite{wei2002} showed that acquiring a new subscriber can be five times harder than retaining an existing one because the latter is less sensitive to both price and sales referrals. Meanwhile, improving subscriber retention can help wireless carriers increase margins \cite{fccreport2009}. Therefore, effective subscriber churn management becomes a priority for many telecom managers as to ensure the sustainable growth of their companies.

Wireless carriers aim at controlling churn through active subscriber retention campaigns. For this purpose, they proactively identify subscribers with high propensity to churn, evaluate the underlying reasons for churn and devise strategies to prevent it. The perplexing nature of churn, however, makes it very difficult to explain and address churn in an efficient and comprehensive manner. Subscribers may churn for many different reasons: they may be unsatisfied with the quality of service; they may get attracted by competitors that provide lower prices; they may decide to acquire a new handset or service that is not currently provided by their carrier. Thus, wireless carriers can hardly provide one single solution to prevent all potential churners from leaving. Therefore, understanding the complexity of the churn problem and disentangling the role of the several factors that can trigger it is fundamental to design sound retention strategies.

In this paper, we look at one such complexity associated with churn. We study the effect of peer influence on churn. If churn is contagious then churn can snowball quickly leading many subscribers to leave the carrier, specially when social networks are dense, as they locally tend to be in the case of wireless services. We apply survival analysis and generalized propensity score to separate peer influence from homophily and measure the effect of the former. We perform our empirical analysis on a massive dataset from a major European wireless carrier that shared call detail records (CDRs) with us. We show how churn increases with friends' churn, a result suggesting that churn managers should prevent group churn instead of looking at churn on an individual basis.

\section{Related Works}
\subsection{Customer Lifetime Value (CLV)}
Many researchers have proposed that looking at the value of existing subscribers is essential because investing in all subscribers will be inefficient \cite{hwang2004}. \cite{rosset2002} introduced the CLV model in the wireless industry to estimate the effect of retention campaigns based on CLV. CLV models help understand subscriber value and provide wireless carriers insights to build cost-effective retention strategies targeted to subscribers with high CLV and low loyalty. \cite{gupta2004} showed that the profit of the wireless carrier is a function of the total subscriber lifetime value. In their theoretical CLV model for the wireless telecommunication industry, the probability of a subscriber to churn is treated as a parameter in the CLV function used to determine how long the subscriber will stay in the network generating future cash flows. Another generalized CLV model proposed by \cite{glady2009} identified churners as subscribers with decreasing CLV. Therefore, the probability of churn is central to the notion of CLV and thus understanding churn is paramount to correctly measure CLV.

\subsection{Churn Prediction}
Today, wireless carriers gather wealthy data deemed useful to perform churn analysis. Numerous data mining techniques have been applied to transform these raw data into useful knowledge. \cite{Hung2006515} described the general framework for this purpose: (i) identify discriminatory features that can differentiate a subscriber with high risk of churn from other loyal subscribers; (ii) extract data from identified features; (iii) select the appropriate data mining techniques to build descriptive or predictive models; (iv) evaluate the performance of these models according to specified criteria, e.g. lift curves. Extensive research has been done on churn prediction in wireless network. Refer to \cite{Ngai20092592} and \cite{verbeke2011} for comprehensive reviews on the application of different algorithms to churn prediction and prevention.

Three disadvantages of pure data mining techniques are worth noting though. First, although many models and algorithms seem to provide satisfactory accuracy in identifying churners, the results obtained dependent not only on method but also on the data used and the features considered by researchers. For example, both \cite{mozer2000} and \cite{hwang2004} used logistic regressions and neural networks to predict churn. The former found that neural networks outperformed logistic regression. However, the latter concluded otherwise. Second, many data mining algorithms are like a ``black box'', which lack interpretability precluding us from understanding the true determinants of subscriber churn. As a result, an agent in a call center might be asked to call a certain subscriber because she is likely to churn. However, very little might have been said to this agent about the underlying reasons that may lead the subscriber to churn, which clearly difficulties the interaction with her. Third, a number of statistical based benchmarking measures of performance for data mining algorithms do not directly yield optimal results in terms of profit maximization from the practitioners' perspective. \cite{Verbeke2012211} proposed a profit-centric performance criterion focusing on the fraction of subscribers that generate the most profit and showed that this approach yielded outcomes different from the ones resting on the best approaches as evaluated by statistically based performance measures.

\subsection{Social Influence}
Advances in studying the effect of social influence on subscriber churn in wireless networks have received much attention in recent times. \cite{dasgupta2008} tried to learn whether the propensity of a subscriber to churn depends on the number of friends that have already churned. This hypothesis is based on premise that a few key individuals may lead to strong ``word-of-mouth'' effects. These individuals may influence their friends to churn, who in turn spread the message to others. So they identified likely churners as those subscribers whose friends have already churned, using a spreading activation-based technique. A set of churners iteratively diffuse the message to other subscribers. Then a subscriber churns once the accumulated level of influence reaches a certain threshold. \cite{dierkes2011} used Markov Logic Networks and propositionalization to develop a predictive model for churn. They also confirmed that ``word-of-mouth'' has a significant impact on subscriber's churn decisions. \cite{Phadke2013} demonstrated that by integrating social factors such as influence from churners into machine learning models can enhance the prediction performance. However, correlation in the behavior among people who share social ties can be explained by both peer influence and their inherent similarities -- homophily \cite{mcpherson2001}. Work that identifies contagious churn separating it from confounding effects such as homophily (friends tend to exhibit similar behavior) is still limited. The contribution of our paper is to analyze contagious churn avoiding misattributing homophily to contagion, or/and vice versa, which typically leads to overestimate the latter. 

Along these lines, \cite{miguel2014} provides a study close to ours, which uses the same dataset to identify peer influence in the the adoption of the iPhone 3G. In their case, they show that peer influence led to roughly 14\% of the observed iPhone adoption during the first year after the introduction of this handset in the market. A number of other studies on the identification of peer influence in the presence of confounding effects in other networked contexts have been proposed\cite{anagnostopoulos2008,bramoulle2009,lafond2010,steglich2004,lewis2012}. For example, \cite{aral2009} used dynamic propensity score matching (PSM) to estimate the effect of contagion in the adoption of an online service by analyzing a community of instant messenger users. Their findings suggest that homophily accounted for much of the adoption previously perceived as peer influence. However, they dichotomize the treatment to use PSM and thus are unable to explore the heterogeneity in treatment levels.  

\section{Data and Descriptive Statistics}

\begin{table*}[htb]
\centering
\begin{tabular}{ll p{4cm}  ccccccccc} \hline
&&   &  \multicolumn{2}{c}{Full Sample}  & & \multicolumn{2}{c}{Churner} & &  \multicolumn{2}{c}{Non-Churner} &      \\ 
&& &  \multicolumn{2}{c}{n=8,345}  & & \multicolumn{2}{c}{n=1,191} & &  \multicolumn{2}{c}{n=7,154} &  \\ \cline{4-5}  \cline{7-8} \cline{10-11}
&& & \multicolumn{1}{c}{Mean} & \multicolumn{1}{c}{Std} & & \multicolumn{1}{c}{Mean} & 
\multicolumn{1}{c}{Std} & & \multicolumn{1}{c}{Mean} & \multicolumn{1}{c}{Std} & t-stat \\

&\multicolumn{1}{l}{Covariates} & \multicolumn{1}{l}{Description}  & \multicolumn{1}{c}{[1]} &  \multicolumn{1}{c}{[2]} & & \multicolumn{1}{c}{[3]} & \multicolumn{1}{c}{[4]} & & \multicolumn{1}{c}{[5]} & \multicolumn{1}{c}{[6]} &   \multicolumn{1}{c}{[7]} \\[.5em] \hline

&\multicolumn{2}{l}{\textit{Monthly usage}} \\
& $n\_calls\_out$ & Calls made & 18.99 & 29.74 & & 6.51 & 12.80 & & 21.07 & 30.89 & $p<0.001$\\ 
& $n\_calls\_in$ & Calls received & 22.96 & 29.45 & & 7.84 & 16.44 & & 25.48 & 30.70 & $p<0.001$\\ 
& $airtime\_out$ & Airtime made & 44.87 & 116.44 & & 13.55 & 40.87 & & 50.08 & 123.88 & $p<0.001$ \\ 
& $airtime\_in$ & Airtime received & 55.54 & 116.63 & & 17.71 & 50.15 & & 61.83 & 123.17 & $p<0.001$ \\ 
& $expenditure$ & Expenditure & 31.78 & 53.36 & & 25.33 & 55.31 & & 32.85 & 52.95 & $p<0.001$\\ [.5em]
& \multicolumn{2}{l}{\textit{Structural properties}} \\
& $frd$ & Number of friends & 8.40 & 9.16 & & 5.76 & 8.91 & & 8.84 & 9.13 & $ p<0.001 $ \\ 
& $\%calls\_out\_other$ & Ratio of calls made to other networks & 0.24 & 0.23 & & 0.27 & 0.27 & & 0.24 & 0.23 & $p<0.001$  \\ 
&  $\%calls\_in\_other$& Ratio of calls received from other networks & 0.25 & 0.23 & & 0.26 & 0.26 & & 0.25 & 0.22 & $p<0.05$ \\ [.5em]
& $tenure$ & Time with EURMO   & 48.10 & 39.32 & & 19.91 & 25.66 & & 52.79 & 39.24 & $p<0.001$ \\[.5em] 
& \multicolumn{2}{l}{\textit{Churner friends}} \\
& $1\mbox{-}call$  $frd\_churn$ & Number of 1-call churner friends  & 1.13 & 2.23 & & 0.81 & 2.05 & & 1.18 & 2.25 & $ p<0.001 $ \\ 
& $3\mbox{-}call$  $frd\_churn$ & Number of 3-call churner friends & 0.56 & 1.25 & & 0.40 & 1.15 & & 0.59 & 1.26 & $ p<0.001 $ \\ 
& $5\mbox{-}call$  $frd\_churn$ & Number of 5-call churner friends & 0.20 & 0.60 & & 0.14 & 0.52 & & 0.21 & 0.61 & $ p<0.001 $ \\[.5em]
\hline
\end{tabular}
\caption{List of covariates extracted from the EURMO network. Descriptive statistics are performed for the our random sample (columns [1] and [2]), churners (columns [3] and [4]) and non-churners (columns [5] and [6]), respectively. Column [7] tests the hypothesis that the means between churners and non-churners are similar.} 
\label{table:covariate}
\end{table*}

We partnered with a major European wireless carrier, hereinafter called EURMO, whig gave us access to its Call Detailed Records (CDRs) between August 2008 and May 2009. For each call we know the caller and the callee, the duration and time of the call and the id of the cell tower used to route the call. Subscribers are identified by their anonymized phone number. For each subscriber, we know their provider and tariff plan at all times. There are roughly 4 million EURMO subscribers in our dataset.

Understanding subscriber churn with prepaid plans is quite different from working with postpaid subscribers. First, we have very limited socio-demographic information on prepaid subscribers. Second, the usage pattern of prepaid subscribers is more irregular than that of postpaid subscribers. Third, prepaid subscribers churn by ceasing usage whereas postpaid subscribers explicitly inform the carrier when they intend to do so. We use the standard in the industry, which is also followed by EURMO and assumed that a prepaid subscriber churns if she places no calls for three of consecutive months.

We use a random sample of 10 thousand subscribers. Two subscribers are called friends if they exchange at least one call in each and every calendar month. We trim from our random sample subscribers with very high degree, which are likely to represent customer service and PBX machines, and with no degree (some subscribers purchase a SIM card but never use it to make calls). We are left with 8,345 subscribers in our sample. We observe network dynamics over time, namely new subscribers join EURMO and existing subscribers leave EURMO every month. Moreover, subscribers call and/or text different friends over time. We aggregate individual subscriber usage and structural properties at the monthly level in our analysis. Table \ref{table:covariate} shows descriptive statistics for covariates used in our study. Over the period of analysis, the subscribers in our sample placed 3.75 million calls and 1,191 of them churned, which amounts to an average monthly churn rate of $2.04\%$. On average, users in our sample have 8.4 friends. Number of friends is the only time-independent covariate used in our study, which we compute across the whole panel of data available to us. This allows us to identify "sticky" friends purging spurious short-lived connections.

We find that churners have much less usage, both in terms of number of calls and airtime, and fewer friends than non-churners. These differences are statistically significant as shown by a t-test in column [7] of this table. Since we know the tariff plan each subscriber holds, we are able to calculate an estimate for monthly expenses (in Euros). Churners contribute with 11\% of the revenues at EURMO. On average, they spend 7.5 euros less per month than non-churners. We observe that all subscribers have much more usage within the network. This is because EURMO operates under a Sender-Pays-All regime and interconnection charges are added to every call, which EURMO passes, partially, to consumers making calls outside the network more expensive.  

We are particularly interested in the association between number of friends who churn and the propensity to churn.  Therefore, we use $n\mbox{-}call$ $frd\_churn$, the friends who churn that exchange at least $n$ calls with the ego in the same calendar month. We find that 343 out of 1,191 (29\%) churners and that 3,197 out of 7,154 (45\%) non-churners have at least one friend that churned during the period of analysis. Table \ref{table:covariate} shows that on average subscribers see 1.13 friends churn, 0.56 and 0.20 friends churn that exchange at least 3 and 5 calls, respectively, with the ego in the same calendar month.

\section{Churn Dynamics}

The panel structure of our data can be analyzed using survival analysis to determine the impact of time-varying covariates on churn \cite{pavel2011}. Survival analysis also allows for controlling for some unobserved individual-level heterogeneity, i.e., some subscribers are more prone to churn for reasons that are not captured in our data (e.g. marketing campaigns by competitors). We employ a Cox Proportional Hazard (PH) model with frailty to estimate the churn hazard rate as:
\begin{table*}[!htb]
\centering

\begin{tabular}{  l l  c  c c c c c  }
  \hline                  
   &  & 1-call & 3-call & 5-call  & 1-call & 3-call  & 5-call   \\
  \multicolumn{2}{l}{Covariates}   &  [1] & [2] & [3] & [4] & [5] & [6]  \\[.5em]
   \hline
   \multicolumn{2}{l}{Influence factor}  & & & & & & \\
  & $frd\_churn$  & 0.375\textsuperscript{***} & 0.215\textsuperscript{***} & 0.458\textsuperscript{***} & 0.369\textsuperscript{***} & 0.333\textsuperscript{***} & 0.449\textsuperscript{***}\\
 &  &  (0.104) &  (0.0787) &  (0.156) & (0.0978) & (0.120) & (0.167) \\[.5em]
 
  \multicolumn{2}{l}{Usage Metrics} & & & & & & \\
   & $n\_calls$  & -0.00646\textsuperscript{***}  &  -0.00621\textsuperscript{***} &  -0.00625\textsuperscript{***} & & & \\
   & &  (0.00159)  & (0.00134) & (0.00148) & & &  \\
   &   $n\_calls^2$  & 0.0000073\textsuperscript{***}  &  0.0000071\textsuperscript{***} &  0.0000072\textsuperscript{***} & & & \\
   & &  (0.0000021)  & (0.0000018) & (0.0000019) & & &  \\
   &  $airtime$ & & & & -0.000829\textsuperscript{**} & -0.000801\textsuperscript{***} & -0.000819\textsuperscript{*}\\
  & &  & & & (0.000382) & (0.000297) & (0.000440)  \\
   &    $airtime^2$ & & & & 0.00000013 & 0.0000001 & 0.00000014\\
  & &  & & & (0.0000001) & (0.0000001) & (0.00000011)  \\
  & $expenditure$ &0.00457\textsuperscript{***} &   0.00488\textsuperscript{***} & 0.00515\textsuperscript{***} & 0.00565\textsuperscript{***} & 0.00554\textsuperscript{***} &  0.00584\textsuperscript{***}\\
  & & (0.00159) & (0.00127) &  (0.00150) & (0.00158) & (0.00120)  & (0.00184)\\[.5em]
  \multicolumn{2}{l}{Structural Metric} & & & & & & \\
  & $frd$ &  -0.0746\textsuperscript{***} &  -0.0659\textsuperscript{***} & -0.0671\textsuperscript{***} & -0.0898\textsuperscript{***} & -0.0803\textsuperscript{***} & -0.0827\textsuperscript{***}  \\ 
  & & (0.0161)  &  (0.0143) & (0.0151) & (0.0150) & (0.0115) & (0.0184) \\[.5em]
  & $month\_dummy$ & Yes & Yes & Yes & Yes & Yes & Yes   \\
   \hline
    \multicolumn{2}{l}{Observations} & 51,488 & 51,488 & 51,488 & 51,488 & 51,488 & 51,488   \\ \hline
   \multicolumn{3}{c}{\textsuperscript{***}$p<0.01$, \textsuperscript{**}$p<0.05$, \textsuperscript{*}$p<0.1$}    
\end{tabular}
\caption{Parameter estimates for Cox PH frailty models on $frd\_churn$ at three thresholds: $1\mbox{-}call$ (columns [1] and [4]), $3\mbox{-}call$ (columns [2] and [5]) and $5\mbox{-}call$ (columns [3] and [6])}
\label{table:hazard}
\end{table*}

\begin{equation*}
h(t,x_i, x_i^{(t-1)}) = \alpha_i h_0(t)\mbox{exp}[\beta x_i + \delta x_i^{(t-1)}]
\end{equation*}

where $h(t, x_i,x_i^{(t-1)})$ is the churn hazard for subscriber $i$ at time $t$, $\alpha_i$ is the Gamma distributed frailty that represents the individual-level random effect,  $h_0(t)$ is the non-paramatric baseline hazard function, $x_i$ and $x_i^{(t-1)}$ include time-independent and time-varying covariates at time $t$, such as usage and friends' churn. We introduce a time-lag on the latter covariates because, similarly to epidemiology cases, there is an induction and latency period between a friend's churn and the ego's churn. Moreover, using lagged covariates obviates simultaneity problems with our estimation. Finally, we also introduce monthly dummies to control for seasonal effects on churn, such as promotions offered by competitors over the summer or during Christmas. 

We count the number of friends who churn that exchange at least $n$ calls with the ego, as explained in the previous section. We vary $n$ in $\{1,3,5\}$ and use $frd\_churn$ to denote the covariate of interest. Table \ref{table:hazard} shows the results obtained. The coefficients on $frd\_churn$ are statistically significant across all model specifications. Note that coefficients for $n=5$ are larger but this can only mildly hint at the fact that churn from stronger friends is more relevant for the ego because standard errors are high. The other covariates behave as expected, in particular, higher expenditure leads to more churn. Yet, we recall that our goal in this paper is not to develop a predictive model of churn. Instead, we are interested in identifying the role of peer influence on churn. The additional covariates in our study are not necessarily used to predict churn. Instead, they are used to characterize the behavior of consumers with respect to how they use cellphone service so that we can, in the next section, compare similar users to try to isolate the effect of peer influence.     

\begin{figure}[!htb]  
  \centering
    \includegraphics[width=0.5\textwidth]{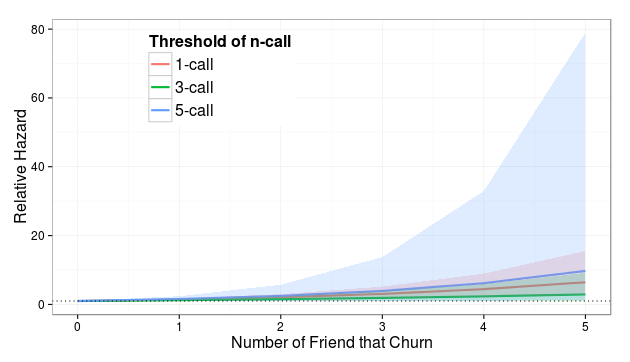}
    \caption{Simulated average churn hazards for models 1-3 in Table \ref{table:hazard}. Ribbons represent 95\% confidence intervals. 10,000 simulations were run per value of $frd\_churn$}
    \label{fig:hazard}
\end{figure}

These results show that the more friends churn the more likely the ego will churn. For example, 0.375 for the first model shown in this table indicates that if one more friend churns then the ego's likelihood of churn vs. no churn increases by $exp(0.375)$ or 45\%. The 0.215 (0.458) in the second (third) model indicates that if one more friend with whom the ego exchanges 3 (5) or more calls in the same calendar month churns then the ego's likelihood of churn vs. no-churn increases by $exp(0.215)$ ($exp(0.458)$) or 24\% (58\%). To better understand the role of friends' churn on the ego's churn we perform Monte Carlo simulations for the relative hazard of churn vs. no-churn using the coefficients and variances of $frd\_churn$ estimated using the six Cox PH models in this table. Figure \ref{fig:hazard} shows how the relative hazard increases with the number of friends that churn.  

\section{Effect of Friends' Churn}

Propensity Score Matching (PSM) is a widely applied method to evaluate the effect of treatments on outcomes of interest \cite{rubin1983, morgan2007} with observational studies. In particular, when the assignment of treatment is not random, selection bias arise because the characteristics of treated and untreated units can differ. The propensity score, defined as the probability of receiving treatment conditional on observed confounding covariates that correlate with both the outcome and treatment, summarizes all the relevant information available to the researcher into a single scalar value. Conditioning on this value, the distributions of the observed characteristics across treated and untreated units become more similar and, therefore, unlikely to drive differences in outcomes. Hence, PSM helps reduce bias. However, PSM still fails to provide full causal interpretations because it does not control for unobserved effects \cite{crs2011}. In any case, and when randomized experiments are unavailable, using PSM increases our confidence when reporting effects, in particular, when one controls for the most important characteristics of the units under analysis. 

Extensions of PSM to cases with continuous treatments \cite{imai2004, imbens2005} have been proposed in the literature under the umbrella of Generalized Propensity Score (GPS). GPS provides a dose-response function (DRF) that measures the relationship between the outcome of interest and the intensity of treatment. In our case, friends' churn (the treatment) is not binary but rather an integer. Egos can be subject to different amounts of treatment as they see more or fewer friends churn. Different treatment intensities can have different effects on the ego's churn (the outcome). Therefore, we use GPS to explore our case of contagious churn in more detail.  

Details of the mechanics beyond GPS can be found in \cite{imbens2005}. For sake of space, we only note the following modeling options: i) the distribution of friends' churn is far from normal, therefore, we followed a parametric solution, as proposed in \cite{Guardabascio2013}, and allow the intensity of treatment to be skewed towards zero, using an exponential functional form whose parameters are estimated by maximum likelihood; ii) we use a polynomial approximation of order two to regress outcomes on treatments and propensity scores, from which we compute the average conditional expectation for the effect of treatment. This polynomial approximation of degree two, with an interaction term between treatment intensity and propensity score, allows for a better understanding of the non-linear effects of treatment than linear regression.  

\subsection{Treatment Exposure and Post-Treatment Periods}

Our dataset spans 10 months of data. However, we need to observe subscribers for 3 months to determine whether they churn. So, in fact, we are limited to a panel with 7 time periods. As discussed earlier, contagious churn is studied after the exposure to friends' churn friend during an induction and latency period. Therefore, we split the period of analysis into two intervals: i) the Treatment Exposure Period (TEP), during which egos observe, and count, their friends churning, represented by $frd\_churn$; ii) the Post-Treatment Period (PTP), during which we observe whether the ego churns. There are several options to define TEP and PTP. However, we need each of these intervals to be sufficiently large. TEP needs to be sufficiently large so that we can observe treatment and, in particular, several intensities of treatment. PTP also needs to be sufficiently large so that we can observe outcomes and, in particular, outcomes triggered by treatment. Therefore, natural choices for TEP and PTP are to include months $\{1,2,3,4\}$ in the former and months $\{5,6,7\}$ in the latter. Another option is to include months $\{1,2,3\}$ in TEP and months $\{4,5,6,7\}$ in PTP. Below we show results using the former definitions. Results for the latter are qualitatively similar and are available upon request.  

\subsection{Balancing Covariates}

We consider 3 levels of treatment intensity: i) T1: at most one friend churns; ii) T2: 2 or 3 friends churn; iii) T3: more than 3 friends churn. We compute a propensity score for each subscriber and for each treatment intensity. Then we test whether the conditional means of the subscribers' covariates given the propensity score are different for subscribers with each treatment intensity and subscribers with other treatment intensities. If the latter are similar then subscribers with different treatment intensities are not different in other aspects of their behavior, which allows us to better associate changes in the outcome (the ego's churn) to changes in the treatment (friends' churn).

The key for any GPS analysis to identify believable effects is to control for the important covariates. Given the richness of our dataset, we are fortunate to observe covariates that, allegedly, capture most of what is important to control for to study contagious churn. First, we control for tenure with EURMO, thus making sure that we take the life cycle of the subscriber within EURMO into account. Otherwise, it would be unreasonable to compare new consumers to old consumers as the latter typically have developed a different level of trust with EURMO, enjoy different prices and, most importably for the issue of churn, experience different switching costs because their lock-in periods are more likely to have expired. Second, we control for number of calls placed and for the percentage of calls to other networks. These covariates together capture well the level of cell usage. Otherwise, it would be unreasonable to compare subscribers with low usage to subscribers with high usage as their engagement with cell phone service is likely very different.

Third, we control for expenditure. While this covariate is highly correlated to usage, controlling for expenditure allows for reducing selection bias introduced by having subscribers choose their tariff plan. Indeed, consumers with different levels of income, and different tariff plans, can pay different amounts of money for the same level of call usage. Such differences could, therefore, be attributed to usage of services other than calling, such as data. Ensuring that both number of calls and expenditure are similar across subscribers allows us to be more confident that we are comparing subscribers that use their cell phone similarly, even for services whose usage we do not observe in our dataset. Finally, we control for the number of friends to make sure that we compare subscribers with similar social circles. Otherwise, it would be unreasonable to compare subscribers that can receive many signals (friends churning) from many friends with subscribers with very few friends can only obtain very limited signals.            

We generate results for $n\mbox{-}call frd\_churn$ for $n=1,3,5$. Table \ref{table:balance} shows how the adjustment by conditioning on the propensity score balances these covariates for the case of $n=1$. We can see that most covariates are different before adjustment but become statistically similar at the 5\% leave after adjustment. Balancing for other values of $n$ and for $(TEP,PTP)=(\{1,2,3\},\{4,5,6,7\})$ yield qualitatively similar results, that is, biases are significantly reduced after adjustment.

\begin{table}[!htb]
\hskip-0.5cm
\begin{tabular}{lcccccccc} \hline
&   \multicolumn{3}{c}{Before adjustment}   & &  \multicolumn{3}{c}{After adjustment} \\
 \cline{2-4}  \cline{6-8}
Covariates   & [T1] & [T2] & [T3]  & & [T1] & [T2] & [T3] \\[.5em] \hline
 $n\_calls$ & \textbf{5.59} & \textbf{-3.21} & \textbf{-2.58} & & 1.85 & -0.77 & -0.94  \\
 $expenditure$ & \textbf{4.36} &  \textbf{-2.68} & \textbf{-2.12} & & 1.42 & -0.64 & -0.77  \\
 $frd$ & \textbf{5.15} & \textbf{-3.27} & \textbf{-2.06} & & 1.15 & -1.24 & 0.27   \\
 $\%call\_other$ & \textbf{2.09} & -0.98 & -1.26 &  &0.84 & -0.36 & -0.83 &   \\
 $tenure$ & \textbf{3.04} & -1.15 & \textbf{-2.04} & &  1.01 & -0.48 & -0.83 &  \\
 \hline
\end{tabular}
\caption{Balance in subscribers' covariates for $1\mbox{-}call$ $frd\_churn$ introduced by conditioning on the generalized propensity score. \textbf{Bold values} indicate covariates
that are different (at the 5\% level) between that treatment intensity and the other treatment intensities.} 
\label{table:balance}
\end{table}

\subsection{The Effect of Treatment}

We use GPS to estimate the conditional mean for the effect of treatment, friends' churn, on the outcome, the ego's churn. We do not report the regression results here, for sake of space and because the estimated coefficients have no direct meaning \cite{imbens2005}. Still, we notice that the coefficients from regressing outcomes on our second degree polynomial on treatment intensities and propensity scores are all statistically different from zero at least at the 10\% level. We estimate the dose response function and derive the marginal treatment effect relative to having no friends that churn. We report marginal treatment effects up to five friends churning, which covers 99.9\% of the observations in our dataset.

\begin{figure}[!htb]  
  \centering
    \includegraphics[width=0.5\textwidth]{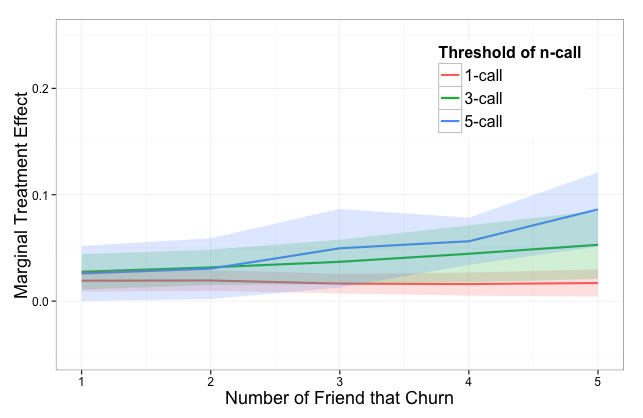}
    \caption{Estimated average marginal treatment effect with TEP:\{1,2,3,4\} and PTP:\{4,5,6,7\} relative to having no friends churning. Ribbons represent the 95\% confidence intervals. Standard errors are obtained via bootstrapping (100 repetitions) }
    \label{fig:teffects}
\end{figure}

Figure \ref{fig:teffects} shows the results obtained for $n=1,3,5$. We observe that having more friends churn increases the likelihood of churn for any $n$ considered in our analysis. This provides evidence of peer influence in churn in wireless networks. Furthermore, we observe that when considering the churn from 3-call and 5-call friends the marginal effect of treatment, that is the effect of having one more of these friends churn, increases with the number of friends that churn. This provides some evidence that churn from stronger friends might be more important. In particular, with GPS, we see that for treatment intensity T3 (that is, more than 3 friends churn), the 95\% confidence interval for 1-call no longer overlaps with the 95\% confidence interval for 5-call, which allows us to claim that the effect of 5-call friends' churn has a large effect than that of 1-call friends' churn when enough friends churn. This is a sensible result showing that enough strong friends churning makes a significant difference on the ego's probability to churn.   

Finally, note that the marginal effect from the GPS analysis is not directly comparable to the relative hazard from the survival analysis. Yet, we can see that the effect of one additional friend churning is at most 10\% in the GPS results. This means that homophily is likely to play a significant role in the correlation of churn across friends that the survival analysis confounds with peer influence. GPS reduces the bias in estimating the effect of peer influence by comparing across similar subscribers. This takes away the effect of all the homophily captured by the covariates that we control for in computing the propensity score.  

\section{Conclusions}
Retaining existing subscribers is of vital importance for wireless carriers to survive in today's dynamic and competitive mobile market. As such, understanding the determinants of churn becomes a priority. Carriers need to confidently identify potential churners to apply appropriate retention strategies aimed at reducing subscriber loss. However, the perplexing and evolving nature of churn still poses significant challenges to churn managers. In this paper, we look at one of such complexities. We examine the effect of peer influence on churn. We do so empirically using a real world dataset of cell phone activity, from which we extract a set of covariates measuring cell phone usage, tenure with the carrier and network structural properties such as number of friends. We analyze a random sample of roughly 10 thousand subscribers. For each subscriber in our sample we define the number of friends that churn with whom the subscriber exchanges 1, 3 or 5 calls in the same calendar month. These different definitions allow for estimating the effect of churn from strong vs. weak friends.

In a preliminary analysis, we fit survival models to our data to correlate the cumulative number of friends that churned up to the previous month and the ego's likelihood of churn in the current month. An additional friend that churns increases the hazard of churn in at least 24\%. In a deeper analysis, we use Generalized Propensity Score (GPS) matching to help reduce the potential selection bias across subscribers in our sample and estimate the effect of contagion. With GPS we can compare subscribers that are similar in a number of relevant covariates and differ only in the intensity of their treatment. The latter is the number of friends that churned. Some subscribers do not have friends that churned while others see as much as 5 friends churn over a period of 7 months. We control for the subscribers tenure with the carrier, cell phone usage, monthly expenditure and the size of the social network. These are the relevant covariates to capture the subscribers behavior with respect to cell phone service and choice of provider. In this framework, we associate differences in the intensity of treatment to differences in the outcome of interest. In our case, the latter is the ego's propensity to churn. 

We find that the ego's propensity to churn increases with the number of friends that churn and in particular with increased numbers of strong friends churning. With GPS, we find that the marginal effect of an additional friend churning is at best 10\%. This means that homophily is likely to drive a significant amount of the correlation between churn and friends' churn that a simple survival analysis is unable to disentangle. With GPS we show that the role of peer influence is still significant in wireless networks: when someone churns from the our carrier, other people may do so due to peer influence. This result, which may well represent a loss of 10\% in revenues for the carrier due to peer influence in churn, suggests that churn managers should consider strategies aimed at preventing group churn instead of looking at subscribers on an individual basis. Finally, we stress that our paper puts together a methodology to measure peer influence in social networks that can be used in other contexts such as learning about the effect of word-of-mouth in the dissemination of new products or services.   

\section{Acknowledgements}
This work was partially supported by the Funda\c{c}\~{a}o para a Ci\r{e}ncia e a Tecnologia (Portuguese Foundation for Science and Technology) through the Carnegie Mellon Portugal Program under Grant SFRH/BD/51153/2010. We are also grateful for the comments and suggestions of three anonymous referees. We thank EURMO and Pavel Krivitsky for providing and organizing data that allowed us to perform this analysis.

%
% The following two commands are all you need in the
% initial runs of your .tex file to
% produce the bibliography for the citations in your paper.
\bibliographystyle{abbrv}
\bibliography{churn_socialcom-Qiwei}  % sigproc.bib is the name of the Bibliography in this case
% You must have a proper ".bib" file
%  and remember to run:
% latex bibtex latex latex
% to resolve all references
%
% ACM needs 'a single self-contained file'!
%
\end{document}